\documentclass[%
 reprint,%
 amssymb, amsmath,%
 aip,apl,%
]{revtex4-1}

\usepackage{docs}%
\usepackage{bm}%
\usepackage[pdftex]{graphicx}
\usepackage{amsmath}    % need for subequations
\usepackage[utf8]{inputenc}
\usepackage{textcomp}
\expandafter\ifx\csname package@font\endcsname\relax\else
 \expandafter\expandafter
 \expandafter\usepackage
 \expandafter\expandafter
 \expandafter{\csname package@font\endcsname}%
\fi
\begin{document}

\title{Long-term transmission of entangled photons from single quantum dot over deployed fiber}%

\author{Zi-Heng Xiang}
\affiliation{Toshiba Research Europe Limited, Cambridge Research Laboratory, 208 Science Park, Milton Road, Cambridge, CB4 0GZ, UK.}
\affiliation{Cavendish Laboratory, JJ Thomson Ave, Cambridge, CB3 0HD, UK.}
\author{Jan Huwer}
\email{jan.huwer@crl.toshiba.co.uk}
\affiliation{Toshiba Research Europe Limited, Cambridge Research Laboratory, 208 Science Park, Milton Road, Cambridge, CB4 0GZ, UK.}
\author{R. Mark Stevenson}
\affiliation{Toshiba Research Europe Limited, Cambridge Research Laboratory, 208 Science Park, Milton Road, Cambridge, CB4 0GZ, UK.}
\author{Joanna Skiba-Szymanska}
\affiliation{Toshiba Research Europe Limited, Cambridge Research Laboratory, 208 Science Park, Milton Road, Cambridge, CB4 0GZ, UK.}
\author{Martin B. Ward}
\affiliation{Toshiba Research Europe Limited, Cambridge Research Laboratory, 208 Science Park, Milton Road, Cambridge, CB4 0GZ, UK.}
\author{Ian Farrer}
\affiliation{Cavendish Laboratory, JJ Thomson Ave, Cambridge, CB3 0HD, UK.}
\affiliation{Present Address: Department of Electronic \& Electrical Engineering, University of Sheffield, Sheffield S1 3JD, UK.}
\author{David A. Ritchie}
\affiliation{Cavendish Laboratory, JJ Thomson Ave, Cambridge, CB3 0HD, UK.}
\affiliation{Toshiba Research Europe Limited, Cambridge Research Laboratory, 208 Science Park, Milton Road, Cambridge, CB4 0GZ, UK.}
\author{Andrew J. Shields}
\affiliation{Toshiba Research Europe Limited, Cambridge Research Laboratory, 208 Science Park, Milton Road, Cambridge, CB4 0GZ, UK.}

\begin{abstract}

Non-classical light sources based on a single quantum emitter are considered as core technology for multiple quantum network architectures. A large variety of sources has been developed, but the generated photons remained far from being utilized in established standard fiber networks.
Here, we report a week-long transmission of polarization-entangled photons from a single InAs/GaAs quantum dot over a metropolitan network fiber. The emitted photons are in the telecommunication O-band, favored for fiber optical communication. We employ a polarization stabilization system overcoming changes of birefringence introduced by 18.23km of installed fiber. Stable transmission of polarization-encoded entanglement with a high fidelity of 91\% is achieved, facilitating the operation of sub-Poissonian quantum light sources over existing fiber networks.

\end{abstract}

\maketitle

With the number of emerging quantum technologies rapidly increasing, the development of quantum networks is becoming more important than ever. Apart from special purpose networks suitable for secure key distribution \cite{ekert1991quantum, qiu2014quantum, tajima2017quantum, sasaki2011field}, more general architectures like a quantum internet\cite{kimble2008quantum, liao2018satellite} are expected to unlock even greater potential for applications like cloud-based quantum computing \cite{cirac1999distributed, lim2005repeat} or quantum sensing \cite{degen2017quantum, medintz2005quantum}. Apart from many other challenges, the most basic discipline in a general purpose quantum network is the generation and detection of entangled quantum bits (qubits). For true scalability not only in distance but also for widespread implementation, approaches are required that enable these two essential tasks in the most simple and robust way. 

\begin{figure*}[th]
\centering
    \includegraphics[width=0.9\textwidth]{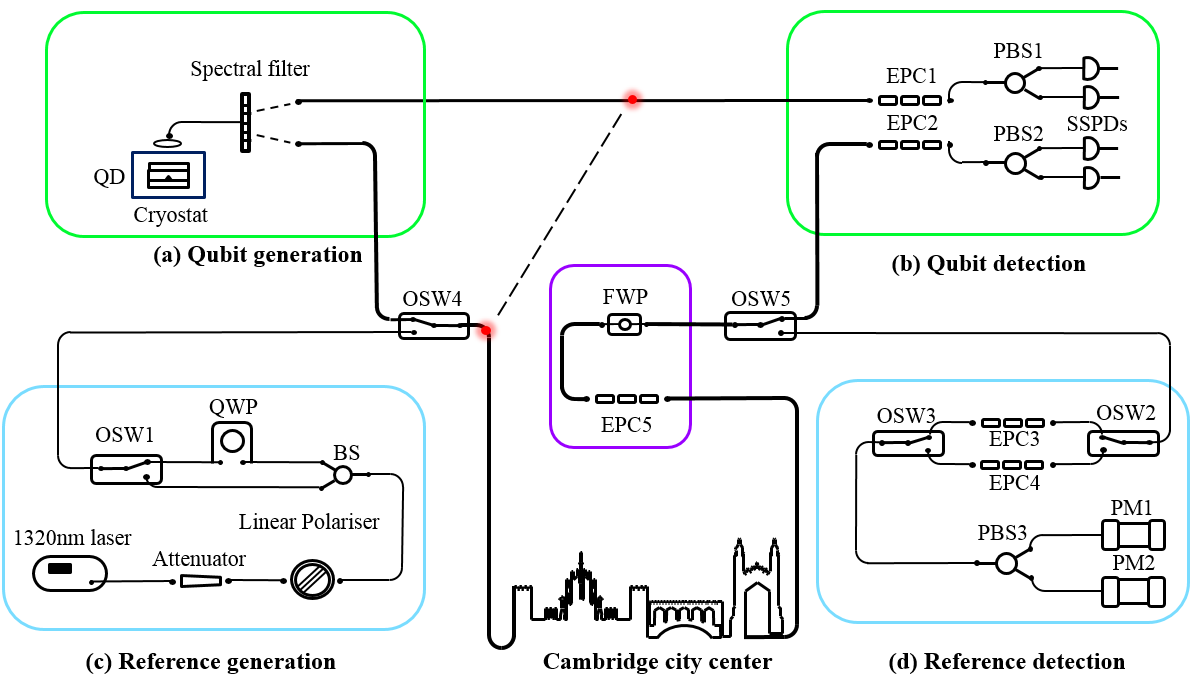}
  \caption{Experimental setup: The green boxes highlight the qubit generation (a) and detection (b) modules. The entangled photon pairs are emitted from a quantum dot (QD). Polarization correlations in different detection bases are measured with a fiber based polarization analyzing setup comprising electronic polarization controllers (EPC) 1 and 2, polarizing beam splitters (PBS) 1 and 2, and 4 superconducting single photon detectors (SSPD). The blue boxes highlight the birefringence stabilization system. Two polarization references are generated from a laser in (c) and detected in (d) in their respective detection bases using EPC 3 and 4, PBS 3 and power meters (PM 1,2). Multiplexing of the two references and the bases is achieved using optical switches (OSW) 1 2 and 3. Detected polarization changes are compensated by applying feedback to a fiber wave plate (FWP) and EPC 5 in the purple square. OSW 4 and 5 are used for multiplexing and de-multiplexing the quantum channel and the reference channel over the field fiber. The spectral filter in (a) and the QWP and linear polariser in (c) are free-space optics, the rest of displayed components are all-fiber based.}
\label{fig:sys}
\end{figure*}

Since the early days, photon pair sources based on spontaneous processes like down conversion \cite{kwiat1995new} and more recently four-wave mixing \cite{li2005optical} were the most prominent choice for photonic entanglement generation. Over the years, the technology has evolved, enabling a large number of quantum-network related experiments. But due to their spontaneous nature, these sources can increase error rates in certain security-relevant applications caused by multi-pair emission \cite{gisin2002quantum}. Sub-Poissonian photon pair sources based on a single quantum emitter such as semiconductor quantum dots (QD) are a promising alternative, as they provide intrinsic security against photon number splitting attacks.

Regarding photonic qubits, there exist two competing approaches for their implementation over optical fiber. Quantum states being encoded in the photon polarization are naturally favored for qubit generation in sources being based on single quantum emitters \cite{benson2000regulated, stevenson2006semiconductor, moreau2001single, wilk2007single, babinec2010diamond}, as they can directly interface with electronic states on optical dipole transitions. However, for long-distance transmission over optical fiber, polarization qubits are affected by random drifts in birefringence due to changing environmental conditions. In contrast, qubits encoded in the phase of subsequent time-bins are realized in a single polarization basis, protecting their information from birefringence-induced rotations and enabling simple stabilization schemes \cite{marcikic2004distribution, yoshino2013maintenance, dynes2016ultra, dixon2015high}. The major disadvantage is that stable interferometers are required for generation and detection, significantly increasing the complexity of the systems.

For certain scenarios, where multiple users are sharing the same quantum channel, the use of polarization encoding might be beneficial. The network provider could take care of the stabilization of birefringence, allowing the end user to operate with cheaper and less sensitive technology, important for scalable and wide-spread implementation. A number of experiments have been performed in the past, employing the transmission of polarization qubits over short specially installed non-telecom fiber without \cite{ursin2004communications} and with \cite{rosenfeld2017event} active stabilization. Efficient feedback systems for the continuous recovery of arbitrary polarization states operating at telecom wavelength have been demonstrated, enabling the stable transmission of polarization qubits from weak coherent sources over long fiber in a laboratory \cite{xavier2008full, peng2007experimental, da2013proof} and from a down-conversion source over deployed fiber  \cite{treiber2009fully}. Here, we implement a similar polarization-control system to stabilize birefringence in 18km of installed fiber across the city of Cambridge. We make use of a telecom-wavelength semiconductor quantum dot as photon source and show stable long-term operation of polarization-entangled photon transmission from a sub-Poissonian emitter over a standard telecommunication-fiber network link.

\begin{figure}[th]
    \includegraphics[width=0.48\textwidth]{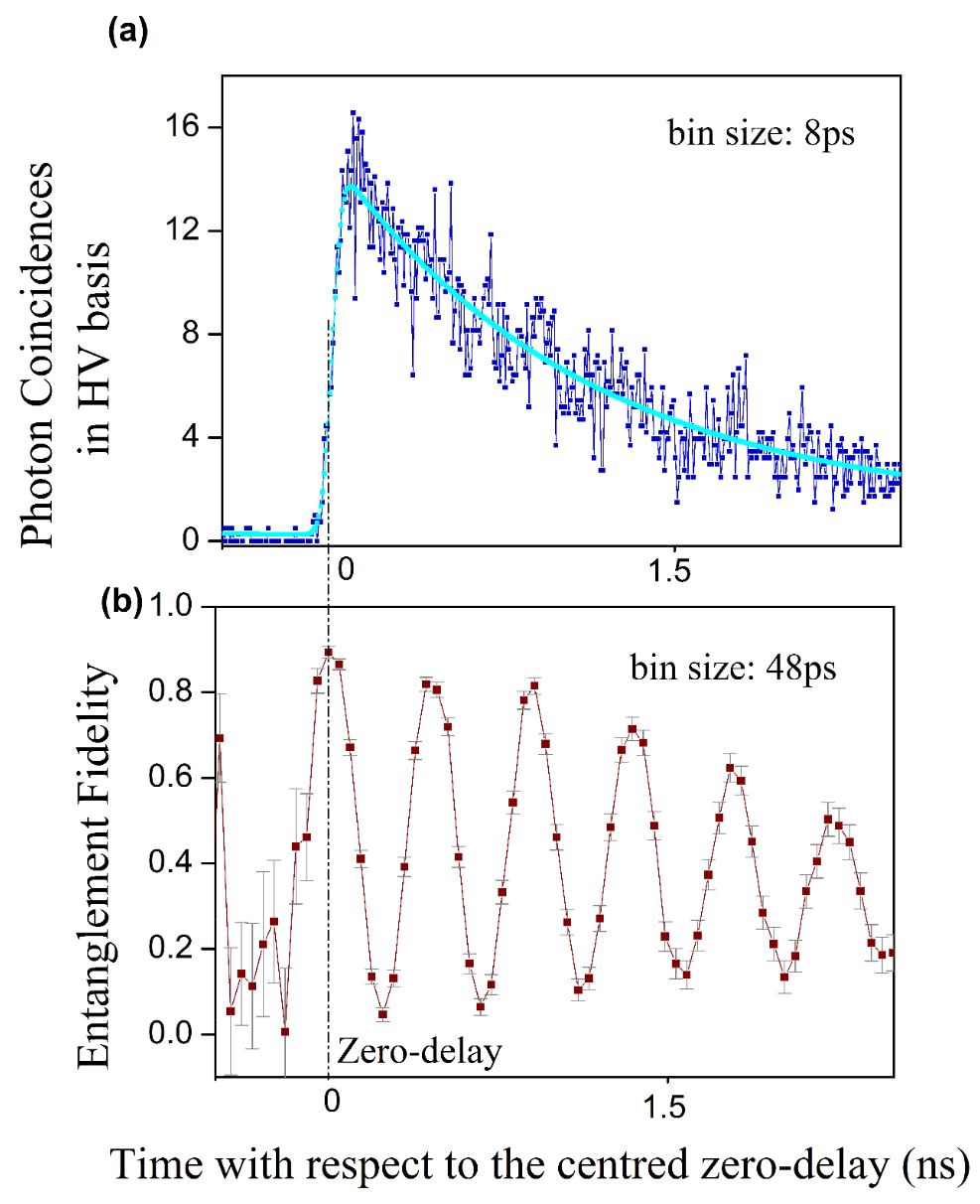}
  \caption{Photon correlations and entanglement fidelity from the QD emitter for a single experimental data set. (a) Normalized coincidences for co-polarized X and XX photons in the HV detection basis. The solid line is an empirical fit to extract the zero delay between the two photons of a pair. (b) Entanglement fidelity as a function of the delay between X and XX photon.}
\label{fig:ent}
\end{figure} 

In recent years, effort has been put to push the emission wavelength of QDs to the standard telecommunication bands, enabling sub-Poissonian entangled photon-pair sources compatible with standard telecommunication infrastructure \cite{alloing2005growth, benyoucef2013telecom, ward2005demand, skiba2017universal}. The InAs/GaAs quantum dot used in this work is located in a PIN structure grown by molecular beam epitaxy with AlGaAs/GaAs stacked Bragg mirrors at the P type and N type layer in order to enhance photon collection \cite{huwer2017quantum}. The ground state of a QD can be occupied with a maximum of two electron-hole pairs which decay in a cascade from the so-called biexciton level (XX) via the intermediate exciton level (X). The two subsequently emitted photons are maximally entangled in their polarization \cite{benson2000regulated} corresponding to the Bell state \(|\Phi^+\rangle = \frac{1}{\sqrt{2}}~(|\textnormal{H}_{\textnormal{X}}\textnormal{H}_{\textnormal{XX}}\rangle + |\textnormal{V}_{\textnormal{X}}\textnormal{V}_{\textnormal{XX}}\rangle)\) with H and V denoting horizontal and vertical polarization respectively. A continuous wave laser at 1064nm is used to optically excite the quantum dot at a temperature of 10K. We apply a bias of -0.165V to the diode, tuning the emission wavelength of XX photons to 1320.0nm and X photons to 1329.4nm.

Figure \ref{fig:sys} illustrates the overall experimental setup. Polarization entangled photon pairs in the telecom O-band are generated from the quantum dot in (a). The light emitted from the device is coupled to single-mode fiber using a confocal microscope configuration, and passes through a free-space spectral filter to isolate entangled qubits into two separate single modes. One photon of a pair is transmitted over short optical fiber to a polarization analyzing setup directly in the lab, whereas the partner photon is sent over a loop-back fiber to the city center of Cambridge before detection back in the laboratory (b). Boxes (c) and (d) display the setup required for polarization stabilization over the deployed fiber.

The main difficulty in transmitting qubits encoded in polarization is to overcome polarization rotations induced by temporal variations of birefringence in installed fibers. The general approach is to inject polarization references which enable efficient detection and compensation of these variations when applying feedback to a set of polarization controllers. Over the network fiber, we observe a strong wavelength dependence of birefringence \cite{keiser2003optical}. This results in a change of polarization states of 20 degrees on average on the Poincar\'{e} sphere for small variations in wavelength of 1nm. Changes in environmental conditions cause a continuous variation of the birefringence \cite{galtarossa2000statistical}, resulting in polarization states of different wavelength evolving in a different random way. Therefore, wavelength-division multiplexing schemes \cite{xavier2008full} could not be implemented for injection and retrieval of the references. We thus apply a time-division multiplexing scheme \cite{treiber2009fully}, with both references and the quantum light at exactly the same wavelength.

For generation of the two references in Figure \ref{fig:sys}(c) we split polarized laser light at 1320nm in two separate spatial modes and use a free-space quarter wave plate (QWP) in one mode for a precise and long-term stable alignment of both polarizations along two orthogonal directions on the Poincar\'{e} sphere. In this configuration, the stabilization of both references efficiently locks arbitrary rotations of the sphere. An optical switch (OSW 1) is used to select either one of the two references. Two standard power meters in (d) are used to evaluate the projection \(\eta\) along the corresponding bases on the  Poincar\'{e} sphere, controlled by electronic polarization controllers (EPC) 3 and 4 and polarizing beam splitter (PBS) 3:
\begin{equation}
\eta = (P_1- P_2)/(P_1+P_2)
\end{equation}
where \(P_1\) and \(P_2\) denote the power values measured by each power meter. OSW 2 and 3 are used for fast and reliable switching between the two detection bases. Once a drop in \(\eta\) is detected for one of the references, EPC 5 and a fiber variable wave plate (FWP) in the purple box are used for applying rotations such that both references are recovered to their original state. Applying a voltage to the FWP results in a clean variable rotation of polarization states around a fixed axis on the Poincar\'{e} sphere. EPC 3 and 4 are set such that one detection basis coincides with the rotation axis of the FWP and the other one is oriented in plane of the rotation. Like this, cross-talk during the recovery of the references is minimized. Both, EPC 5 and FWP are all-fiber based standard components featuring low insertion loss(\(< 0.8\)dB) and low-voltage operation.

\begin{figure*}[th]
  \centering
    \includegraphics[width=\textwidth]{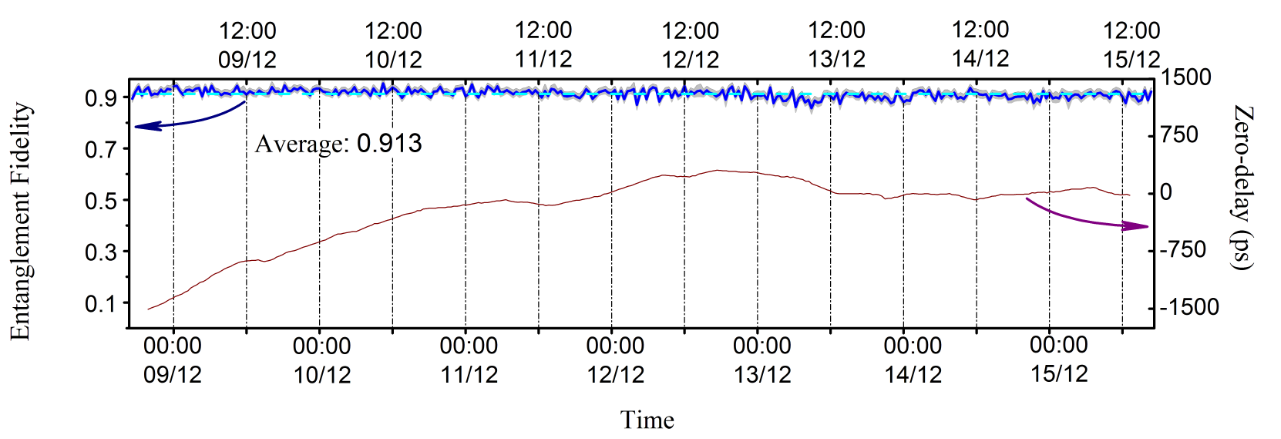}
  \caption{Entanglement fidelity and relative change of qubit transit time over the field fiber for 7 days of continuous operation. The mean fidelity is \((91.3 \pm 1.4)\%\). The gray shaded area indicates one standard deviation.  Error bars for the delay values are negligible.}
\label{fig:fid}
\end{figure*}

Photon entanglement is analysed by processing time-resolved correlations $c_{\textnormal{MN}}$ between measured X and XX photon arrival times for co- and cross-polarized states M and N in the three detection bases HV, DA (diagonal, anti-diagonal) and RL (right-, left-circular). These bases are initially calibrated using EPC 1 and 2 after injecting polarization references at the position of the QD, not displayed in Figure \ref{fig:sys}. The fidelity to the maximally entangled Bell  \(|\Phi^+\rangle\) state is calculated as \cite{michler2009single}
\begin{equation} 
F =  ( 1+C_{\textnormal{HV}}+C_{\textnormal{DA}}-C_{\textnormal{RL}})/4 
\end{equation}
with $C_{\textnormal{MN}} = (c_{\textnormal{MM}}-c_{\textnormal{MN}})/(c_{\textnormal{MM}}+c_{\textnormal{MN}})$ being the correlation contrast. Both, source and superconducting single photon detectors are running continuously during the measurements. The detection basis is switched every 10\,min such that the fidelity can be evaluated for every 30\,min of correlation data. An example for one set of data is shown in Figure \ref{fig:ent} (b). Due to imperfections in the QD morphology, the fidelity follows a time-dependent oscillation caused by interference of two non-degenerate decay channels in the cascade introduced by the so-called fine structure splitting \cite{stevenson2008evolution, ward2014coherent}. In order to extract the entanglement fidelity to  \(|\Phi^+\rangle\), a post selection window is applied to each data set, isolating  correlations in a 48ps interval around the zero delay of the cascade. As changes in the ambient temperature result in a change of the effective optical length of the installed fiber, an exact knowledge of the zero delay is crucial in this analysis. As both photons of a pair are detected by the same single photon counting unit, no active synchronization is required. We extract the time-of flight for each 30min chunk of data from a fit to correlations $c_{\textnormal{HH}}$ between X and XX photons measured in the HV polarization basis (Figure  \ref{fig:ent} (a)). Over this time scale the drift is typically negligible. The overall timing jitter of the photon detection system is 70ps and correlation data is analyzed on a 48ps timing grid.

\begin{figure}[bp]
\centering
\includegraphics[width=0.48\textwidth]{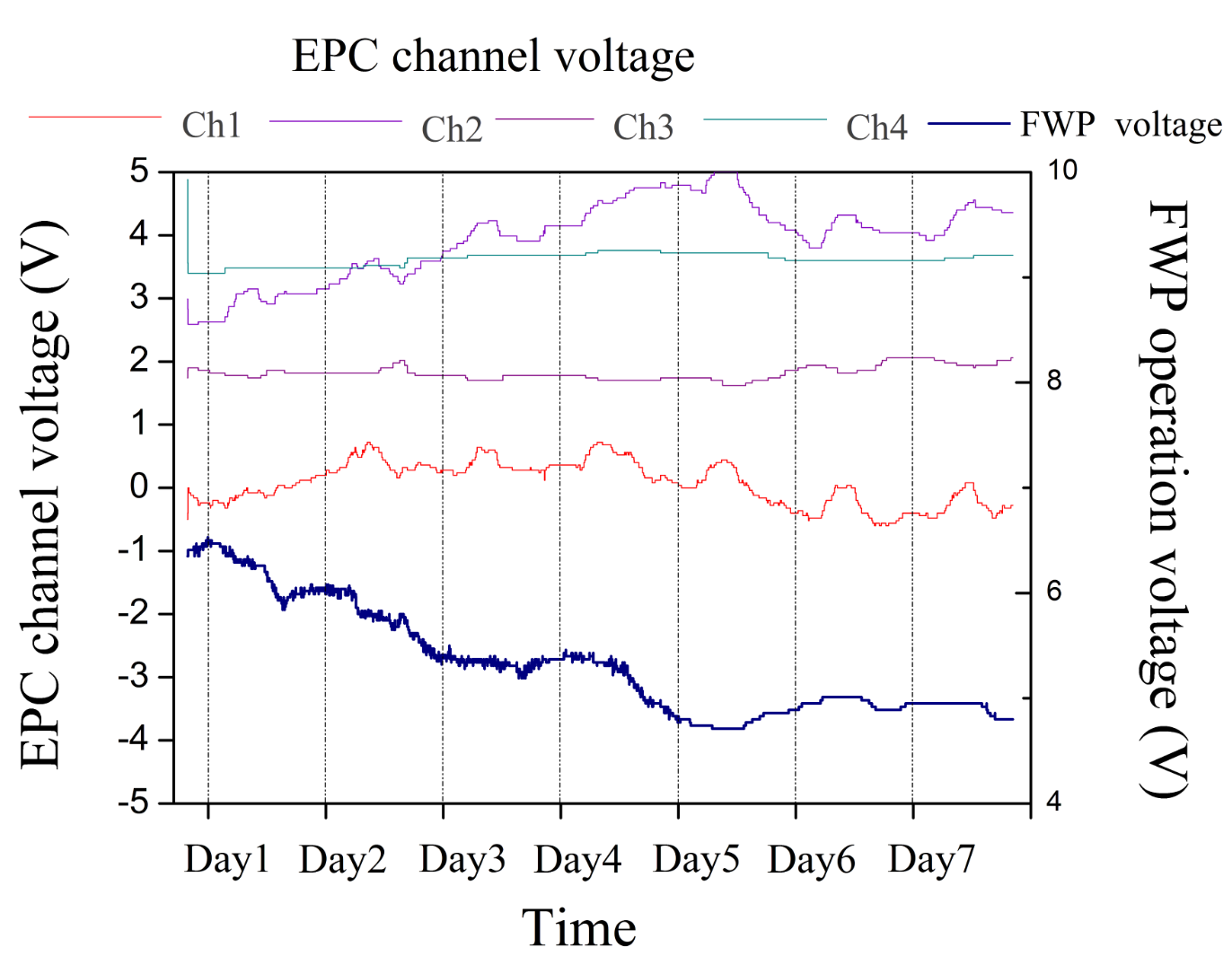}
\caption{Operation voltage for EPC5 and FWP over the course of the measurement. }
\label{fig:ope}
\end{figure}

The entanglement fidelity has been continuously recorded for 7 days, with the XX photons being sent over the field fiber. The overall results are shown in Figure \ref{fig:fid}. Displayed is the entanglement fidelity and the change in time-of-flight as a function of measurement time. A constantly high entanglement fidelity is achieved over the entire week, with an average value of  \((91.3 \pm 1.4)\%\). For comparison, the entanglement fidelity measured without sending the photons over the loop-back link is  \((94.7 \pm 1.7)\%\). During the course of the week, rain has dropped and snow has fallen over Cambridge, with the temperature variation from -4 to 7 \textdegree C. The impact of these changing environmental conditions can be seen in the drift of the time-of-flight of photons of 1.82ns over the first 4 days. Figure \ref{fig:ope} shows the voltages applied to the four channels of the EPC and the FWP as a result of the feedback generated to keep the birefringence stable. Superimposed to the steady drift over the first 4 days, one can see clear oscillations in EPC and FWP operation voltages corresponding to the day-night cycle of temperatures.

When using time-division multiplexing schemes, it is essential to achieve high duty cycles for qubit transmission. Over the course of the measurement, the polarization references were checked for 0.5s alternatingly every 60s, corresponding to a duty cycle of 99\%. Once the projection value \(\eta\) dropped below a threshold of 98.5\%, feedback for recovery was applied which took around 5s on average, resulting in a duty cycle of 92\%. Apart from the threshold-based check, the recovery system was activated every 11 minutes to maximize the alignment for both references. Thus, an overall duty cycle of 98\% with polarization maintenance over 98.5\% was realized for the qubit transmission. Since the optical switches used for multiplexing allow for switching speeds above 100Hz it will be straight forward to adopt the system to field fibers experiencing much higher drift rates than the link available for this work.

Quantum light sources based on single quantum emitters are one of the most precious resources in a quantum network. As the brightness of these sources cannot be easily tuned like attenuated lasers, it is crucial that independent feedback systems add as little loss as possible. With this in mind, the system has been designed using standard fiber components with low insertion loss. The combined losses of optical switches OSW 4 and 5 used for multiplexing between quantum signal and references, and EPC 5 and FWP used for stabilization, add up to 3.49dB, which can be further reduced by splicing. The installed loop-back fiber has a total length of 18.23km and a measured loss of 11.70dB for the transmission of photons at 1320nm.

In summary, we have reported for the first time the long-term transmission of polarization qubits from a single quantum-dot emitter over 18.23km of installed standard telecom fiber. The transmitted photons exhibit a constantly high entanglement fidelity of 91.3\% with their partner photons measured locally, corresponding to a drop in fidelity by only 3.4\% with respect to the source properties. The system has a low systematic loss and a high duty cycle that allows a high transmission efficiency of the qubits. The results demonstrate that quantum-dot emitters natively operating at telecom wavelength, combined with the deployment of entangled qubits over installed fiber, provide a reliable and stable technology which is highly competitive in terms of its low level of complexity, regarding qubit generation and robust detection schemes.

\begin{acknowledgments}
The authors acknowledge partial financial support from the Engineering and Physical Sciences Research Council (EPSRC) and the EPSRC Quantum Technology Hubs in Quantum Communications. Z-H. Xiang acknowledges support from the Cambridge Trust and China Scholarship Council(CSC). The authors would like to thank J. Dynes for technical discussions and A. Wonfor for arrangement of the loop-back fiber link.
\end{acknowledgments}

\end{document}